\documentclass[twocolumn,showpacs,preprintnumbers,amsmath]{revtex4}
\usepackage{dcolumn}
\usepackage{bm}
\usepackage{graphicx}
\usepackage{bm}
\begin{document}
\title{Probing a ferromagnetic critical regime using nonlinear susceptibility}
\author{Sunil Nair and A. Banerjee}
\affiliation{Inter University Consortium for D.A.E. Faclilites\\ University Campus, Khandwa Road. Indore. INDIA 452 017.}
\begin{abstract}
The second order para$-$ferromagnetic phase transition in a series of amorphous alloys  $(Fe{_5}Co{_{50}}Ni{_{17-x}}Cr{_x}B{_{16}}Si{_{12}})$ is investigated using nonlinear susceptibility.  A simple molecular field treatment for the critical region shows that the third order suceptibility $(\chi{_3})$ diverges on both sides of the transition temperature, and changes sign at $T{_C}$. This critical behaviour is observed experimentally in this series of amorphous ferromagnets, and the related assymptotic critical exponents are calculated.  It is shown that using the proper scaling equations, all the exponents necessary for a complete characterization of the phase transition can be determined using linear and nonlinear susceptiblity measurements alone.  Using meticulous nonlinear susceptibility measurements, it is shown that at times $\chi{_3}$ can be more sensitive than the linear susceptibility $(\chi{_1})$ in unravelling the magnetism of ferromagnetic spin systems. A new technique for accurately determining $T{_C}$ is discussed, which makes use of the functional form of $\chi{_3}$ in the critical region. 
\end{abstract}
\maketitle
\section{Introduction:} 
Nonlinear effects become more pronounced in the vicinity of a phase transition, and the magnetisation (M) measured in the presence of an exciting field (H) can be written as 
\begin{eqnarray}M = M{_0} + \chi{_1}H + \chi{_2}H{^2} + \chi{_3}H{^3} +......
\end{eqnarray}
\\Where $\chi{_1}$, $\chi{_2}$ and $\chi{_3}$  are the first, second and third order susceptibilities respectively.  The higher order susceptibilities contain a wealth of information, but owing to the fact that they are usually a couple of orders smaller in magnitude than the first order susceptibility, are correspondingly more difficult to measure.\newline  The second order susceptibility ($\chi{_2}$) arises due to the presence of a symmetry breaking field and is observed only in the presence of a DC field, or in materials with a large permanent magnetisation. Though relatively unexplored, $\chi{_2}$ has been used to show the coexistence of  spin glass phase and ferromagnetic order in a reentrant spin glass system\cite{rang}.\\ The third order susceptibility ($\chi{_3}$) has been commonly employed to characterise the spin glass transition, where a negative cusp in $\chi{_3}$ was used to identify the onset of a spin glass transition and was used as a direct probe of the divergence of the Edward-Anderson order parameter\cite{bind}. $\chi{_3}$ has also been used to distinguish a spin glass from a superparamagnet\cite{ash1,ash2,ash3}.  Among long range order systems, though $\chi{_3}$ deduced from dc magnetisation measurements has been used for studying the effect of quadrupolar interactions in rare earth compounds\cite{morin}, it is interesting to note that nonlinear susceptibility measurements on ferromagnetic\cite{bitoh,sato} and antiferromagnetic\cite{ram} materials are not common, and very few reports exist in literature. Moreover , most of the previous experimental works on nonlinear susceptibility in long range order systems are predominantly qualitative and not quantitative in nature.  As far as the study of phase transitions is concerned, though the second order paramagnetic to ferromagnetic phase transition has been extensively studied using linear susceptibility\cite{kaul}; the behavior of $\chi{_3}$ across the transition has been relatively less explored.  \\To bring out the advantages as well as the limitations of nonlinear susceptibility measurements in the investigation of a para-ferromagnetic phase transition, we present a comprehensive study on a set of Fe-Co-Ni-Cr-B-Si amorphous alloys.  These set of alloys are chosen because they were well studied in the past, and it was found through linear ac susceptibility and high field dc magnetisation measurements that they are reasonably well behaved 3-D Heisenberg ferromagnets\cite{akm,akm1,akm2}.  As such, amorphous ferromagnets have been extensively studied using linear susceptibility to understand the influence of quenched disorder on the critical behaviour of spin systems.  The effect of disorder as seen in the values of the critical exponents and the width of the assymptotic critical regime have been studied and reviewed extensively\cite{kaul1}. Nonlinear susceptibility as measured in an amorphous ferromagnet has been reported before\cite{chi}, but to the best of our knowledge, this is the first study on atleast this class of materials, where linear and nonlinear susceptibility alone is used for a complete characterisation of the critical regime. \\ At the outset, we present a simple molecular field treatment to show that $\chi{_3}$ diverges on both sides of the transition, and changes sign at the transition temperature.  Thus the criticality is observed experimentally, and the related  asymptotic critical exponents are calculated.  It has been shown that using the relevant scaling equations, all the necessary exponents, required for a complete characterization of the phase transition can be determined using linear and nonlinear susceptibility measurements alone.  This can be considered as a major advantage of measurement of nonlinear susceptibility.  Also, owing to the functional form of $\chi{_3}$ across the phase transition, the determination of the true (zero field)  transition temperature ($T{_C}$) can be made to a better accuracy.  It is shown that subtle features not discernable using normal linear susceptibility measurements, can be picked up using nonlinear susceptibility, making it a very useful tool for studies of spin systems in the critical region. Hence, in this paper, we show the following advantages of $\chi{_3}$:\\(i) Determination of all the relevant critical exponents\\(ii) Accurate determination of $T{_C}$\\(iii)Detecting subtle features not seen in linear susceptibility measurements.\\ It is important to note that though the present study deals with amorphous ferromagnets, criticality in $\chi{_3}$ can be used equally effectively in the study of phase transitions in their crystalline counterparts. 
\maketitle\section{Experimental details}
A well characterised and extensively studied series of amorphous ferromagnets ($Fe{_5}Co{_{50}}Ni{_{17-x}}Cr{_x}B{_{16}}Si{_{12}}$) with x = 5, 10, 15  referred to as A2, A3 and A4 have been used in this study.  These samples  prepared by melt quenching are in the form of ribbons of approximate dimensions (10mm X 1mm X 0.03mm).\cite{akm}  \newline In one of the earlier works\cite{morin}, $\chi{_3}$ was determined by fitting the measured isothermal magnetisation using the equation 1, or alternatively by plotting M/H vs $H{^2}$, and calculating the slope in the linear low field range. \newline Another technique involves using the Arrots plots, $M{^2}$ vs $H/M$ , where the slope in the low field range can be related to $\chi{_3}$, through the relation 
\begin{eqnarray}\chi{_3} = - \chi{_1}{^4} (dM{^2}/d(H/M))
\end{eqnarray}.
\newline However, these methods are unsuitable for studying the critical behaviour across the ferromagnetic transition primary due to two reasons; \newline(i) The measuring fields are relatively higher, and this can at times mask the true critical behaviour of the system \newline(ii)$\chi{_3}$ is determined after fitting the dc magnetisation data.  This method is not only time consuming, but also introduces significant errors in the value of $\chi{_3}$. \newline In the present study the linear and nonlinear susceptibilities are measured by monitoring the change in the induced voltage across two oppositely wound secondary coils in a home made susceptometer\cite{ashna}. Both the linear and nonlinear susceptibility was determined by using a Lock-in-Amplifier to measure the signal seen at $\omega$ and 3 $\omega$ respectively, where $\omega$ is the exciting frequency. Neglecting the higher order terms, the voltage measured at $3\omega$ is given by 
\[
 V{_3}{_\omega} = \frac{3}{4} \omega\chi{_3}H{^3}
\]
where H is the magnitude of the applied field. \newline It is normal practice to bundle a few ribbons together to increase the measured signal, but in this case, a solitary ribbon was stuck on to a sapphire single crystal, which made up the sample holder, with the temperature sensor and a non magnetic heater on the same crystal. This was done to avoid temperature gradients across the samples.  A calibrated Platinum resistance thermometer (PRT) was used for measuring the temperature and control to an accuracy of 0.01K was achieved using a Lakeshore temperature controller DRC-93A.  Exciting fields from about 100 mOe to a few Oe and frequencies ranging from 73 Hz to 1.33kHz were used for different experimental runs.  All the measurements were performed with the applied field parallel to the long edges of the samples to minimise the effect of the demagnetisation fields.
\maketitle\section{Results and Discussions} 
\subsection{Linear Susceptibility}
AC susceptibility is amongst the finest tools for the study of critical phenomenon, not only due to the relative ease with which measurements can be done but also due to the fact that very low field measurements can be performed.  The most important advantage is that it directly gives the true initial susceptibility ($\chi{_0}$), which otherwise has to be calculated from data taken at comparatively higher measuring fields.  Considering that fact that the ideal second order paramagnetic-ferromagnetic phase transition is defined in zero external field, ac susceptibility becomes a very good tool for studying criticality, as larger exciting fields used during the course of measurements are likely to mask the true critical behaviour of the system under study.  Also, the Kouvel Fisher [K-F] analysis\cite{kf} of the ac susceptibility data is a good means of accurately determining both the transition temperature ($T{_C}$) as well as the susceptibility exponent ($\gamma $).  Figures 1a, 1b and 1c show the Kouvel Fisher plots ($1/\chi{_0}d/dt(\chi{_0}{^{-1}}$) Vs T)  for the three samples.  The inverse of the slopes directly gives us the value of the susceptibility exponent $\gamma$  and the intercepts of the straight line provides us with an independent estimate of the transition temperature $T{_C}$.  This series has been well studied in the past\cite{akm}, and the values of $T{_C}$ and $\gamma$  reported are in reasonable agreement with the values we have obtained, as is shown in Table 1. 
\begin{table}
\caption{\label{tab:table 1}Values of the susceptibility exponent ($\gamma$) and the transition temperature $T{_C}$ as determined from the Kouvel-Fischer analysis of the first order susceptibility.  The values determined by the earlier workers is given for the sake of comparison.}
\begin{ruledtabular}
\begin{tabular}{cccc}
&A2 &A3 &A4\\
\hline
$T{_C}$ &  267.44 & 222.76 & 174.37 \\
&267$\footnotemark[1]$ & 222.2$\footnotemark[1]$ & 174$\footnotemark[2]$\\
\hline
$\gamma$ & $1.16\pm 0.008$ & $1.388\pm0.01$ &$1.41\pm0.01$\\
&1.19$\footnotemark[1]$ &1.38$\footnotemark[1]$ &1.73$\footnotemark[2]$\\
\end{tabular}
\end{ruledtabular}
\footnotetext[1]{Values determined in ref \cite{akm}using ac-susceptibility measurements}
\footnotetext[2]{Values determined in ref\cite{akm}using high field dc magnetisation }
\end{table}
\subsection{Nonlinear Susceptibility}
\subsubsection{Theory}
The critical behaviour of $\chi{_3}$  has been investigated in the past using the Sherrington-Kirkpatric model\cite{wada} and the Bethe approximation\cite{fuj}.  The behaviour of $\chi{_3}$ was shown by a simple molecular field theory\cite{sato} by T. Sato and Y. Miyako.  Here we present a detailed calculation based on the molecular field approach.\newline Consider a collection of n particles, each with spin $1/2$. When an external field H is applied, the net magnetic moment is given by 
\begin{equation} 
m = n\mu\tanh(\frac{\mu H}{kT})
\end{equation} 
\newline where $\mu$ is the Bohr's Magneton and k is the Boltzmans constant. The net applied field which the dipole sees is given by 
\[
H = H +\lambda M
\]
where $\lambda$ is the molecular field constant. Thus, the magnetisation can now be written as  
\begin{equation} 
M = N\mu\tanh \frac{\mu}{kT}(H+\lambda M)).
\end{equation}
where $N$ is the particle density ($N = n/V$).
\newline
\newline{\bf  Case A:} $T> T{_C}$
\newline In the paramagnetic region, as the magnetisation (M) has inversion symmetry with respect to the applied field H, the magnetisation can be written as 
\begin{equation} 
M = \chi{_1}H + \chi{_3}H{^3} + \chi{_5}H{^5} + .... 
\end{equation}
Substituting this  in equation 4, we get 
\begin{equation} 
\chi{_1}H +\chi{_3}H{^3} = N\mu\tanh\frac {\mu} {kT}(H+\lambda[\chi{_1}H+\chi{_3}H{^3}+...])
\end{equation}
\newline Now, using the expansion 
\[
\tanh(x) = x -\frac{1}{3}x{^3};
\] 
\newline using  the expression for $T{_C}$ from the mean field theory,  $kT{_C} = N\mu{^2}\lambda$ 
\newline and comparing the co efficients of H, $H{^3}$, etc we get 
\begin{equation} 
\chi{_1} = \frac {N\mu{^2}}{kT{_C}}\frac{1}{(\frac{T}{T{_C}}-1)}
\end{equation} 
\begin{equation}
\chi{_3} = -\frac{N\mu{^4}}{3k{^3}T{^3}}\frac{T}{T{_C}}\frac{1}{(\frac{T}{T{_C}}-1){^4}}
\end{equation} 
It is clear that in the limit $T\Rightarrow T{_C}{^+}$, $\chi{_1}$ diverges positively, whereas $\chi{_3}$ shows a negative divergence.\newline
\newline{\bf Case B:} $T< T{_C}$
\newline In the ferromagnetic region, the emergence of spontaneous magnetisation causes M to loose its inversion symmetry with respect to H, and now the magnetisation can be written as  
\begin{equation} 
M = M{_0} + \chi{_1}H + \chi{_2}H{^2} + \chi{_3}H{^3} + ....
\end{equation}  
where $M{_0}$ is the spontaneous magnetisation. Substituting this in equation 4 we get 
\begin{equation} 
M = N\mu\tanh[\frac{\mu}{kT}(\lambda M{_0} + (\lambda \chi{_1}+1)H + \lambda \chi{_2}H{^2} + \lambda \chi{_3}H{^3} )]
\end{equation} 
Expanding $\tanh(x)$ and comparing the coefficients of the different powers of H, we get 
\begin{equation}
M{_0} = N\mu[\frac{\mu\lambda M{_0}}{kT} - \frac{1}{3}\frac{\mu{^3}\lambda{^3}M{_0}{^3}}{k{^3}T{^3}}]
\end{equation}
\begin{equation}
\chi{_1} = \frac{N\mu{^2}}{kT{_C}}\frac{[1-(\frac{T{_C}}{T}){^2}(\frac{M{_0}}{N\mu}){^2}]}{ [\frac{T}{T{_C}}-(1-(\frac{T{_C}}{T}){^2}(\frac{M{_0}}{N\mu}){^2})]}
\end{equation}
\begin{equation}
\chi{_2} = -\frac{kT{_C}M{_0}\chi{_1}{^3}}{N{^3}\mu{^4}}\frac{1}{[1-(\frac{T{_C}}{T}){^2}(\frac{M{_0}}{N\mu}){^2}]{^3}}
\end{equation} 
\begin{widetext}
\begin{equation}
\chi{_3} = \frac{\chi{_1}{^4}}{[1-(\frac{T{_C}}{T}){^2}(\frac{M{_0}}{N\mu}){^2}]{^4}}(\frac{\lambda}{N{^2}\mu{^2}})[2\frac{{\lambda}T{_C}}{T}\frac{M{_0}{^2}}{(N\mu){^2}}\frac{\chi{_1}}{[1-(\frac{T{_C}}{T}){^2}(\frac{M{_0}}{N\mu}){^2}]} - \frac{1}{3}\frac{T}{T{_C}}]
\end{equation}
\end{widetext} 
Equation 11 is solved for the spontaneous magnetisation and yields for $T< T{_C}$ 
\begin{equation}
(\frac{M{_0}}{N\mu}){^2} = 3(\frac{T}{T{_C}}){^2}(1-\frac{T}{T{_C}})
\end{equation}
\newline Substituting this value of $(\frac{M{_0}}{N\mu}){^2}$ in the value of $\chi{_1}$ and simplifying, we get 
\begin{equation}
\chi{_1} = \frac{N \mu{^2}}{2kT{_C}}\frac{[3\frac{T}{T{_C}}-2]}{[1-\frac{T}{T{_C}}]}
\end{equation} 
\newline Hence it is clear that  $\chi{_1}$ has a positive divergence as $T\Rightarrow T{_C}{^-}$. \newline Now, substituting equations 15 and 16 in the value of $\chi{_3}$, and simplifying, we obtain 
\begin{equation}
\chi{_3}=\frac{8}{3}\frac{(\frac{N\mu{^2}}{2kT{_C}}){^4}}{(1-\frac{T}{T{_C}}){^4}}\frac{\lambda T}{(N\mu){^2} T{_C}}
\end{equation}
\newline where it is obvious that like $\chi{_1}$, $\chi{_3}$ too diverges positively in the limit $T\Rightarrow T{_C}{^-}$ 
\subsubsection{Experiments} 
Normally, $\chi{_3}$ is seen to show a sharp negative peak around the transition.  However, as is clear from the previous section, theroretically, $\chi{_3}$ is expected to diverge on both sides of the phase transion, and change sign exactly at $T{_C}$. However, this critical behaviour is not easy to observe.  The reason for it is that the expansion of M in terms of higher powers of H, which was shown in the previous section is truely speaking only valid for small values of H. Hence, to observe the true critical behaviour, measurements have to be done at very low measuring fields, and higher fields tend to smear off the transition. It is difficult to predict  apriori the fields at which this critical behaviour will be seen, as it depends on the extent of nonlinearity in the system, and thus differs from one sample to another. It must be kept in mind that at higher measuring fields, the contribution from domains can become large, thus masking the true critical behaviour of the spin system.\newline Careful low field measurements have shown this critical feature in this series of samples. Fig 2a, 2b and 2c  shows the critical behaviour in the samples A2, A3 and A4 respectively.  As predicted in the previous section, $\chi{_3}$ changes sign across $T{_C}$ and the exact $T{_C}$ can be directly determined from the crossover of $\chi{_3}$ in the temperature axis.  This experimental $T{_C}$ values are given in Table 2.  It is to be noted here that the $T{_C}$ determined from the crossover of $\chi{_3}$ matches well with that determined from the Kouvel-Fischer plot (Table 1). This matching of $T{_C}$'s from two different measurements not only substantiate the fact that the measured $\chi{_3}$ is a genuine response of the spin system, but also that the crossover of $\chi{_3}$ can be used as a direct method to determine the $T{_C}$.  The other advantages of determination of $T{_C}$ from the crossover of $\chi{_3}$ will be discussed in the next section.\newline  As is seen in fig 2a, a double transition showing two crossovers is observed in the sample A2. As will be shown later, this sample gives an uncharacteristically low value of the susceptibility exponent($\gamma$), a fact observed by the earlier workers as well.  Considering the fact that this sample has the largest percentage of Ni and is closest to the critical concentration$(x{_c})$, this double transition is not difficult to explain, as in this class of materials, the formation of clusters showing a distribution of $T{_C}$'s have been speculated\cite{akm,kaul1}.  However, it is interesting to note that no direct evidence of such a distribution is evident from the linear susceptibility measurements.  This only goes on to show that low field nonlinear susceptibility is a more sensitive tool than the linear susceptibility, as far as studies of spin systems near the transition is concerned. \newline As mentioned earlier, the field in which the measurement is performed, is an important consideration in the study of criticality in $\chi{_3}$.  This is clearly shown in fig 3, which shows the field dependence of $\chi{_3}$ in the sample A3 at a measuring frequency of 133.33 Hz.  It is clearly seen that the critical feature is sharpest at the lowest measured field, and increasing the measuring field wipes off this critical behaviour. It is interesting to note that we have observed a non trivial frequency dependence  in this series of samples . This could possibly be due to the presence of superparamagnetic clusters, as no frequency dependence would be expected in a long range ferromagnetically ordered system.  These results will be dealt with in a later communication.\newline Large uncertainties in $\frac{d}{dT}(\chi{_3}{^{-1}})$ make it very difficult to determine the value of the critical exponent associated with the third order susceptibility ($\gamma{_3}$), using the Kouvel-Fischer formalism.  Hence the value of $\gamma{_3}$ is determined by plotting a double logarithmic plot of $(-\frac{3}{4})\chi{_3}H{^2}$ Vs $\epsilon$  where $\epsilon = (\frac{T-T{_C}}{T{_C}})$ is calculated by taking the crossover point of $\chi{_3}$ as $T{_C}$. Fig 4a, 4b and 4c shows the double logarithmic plots for samples  A2, A3 and A4 respectively.  As shown in the figure, as $T \rightarrow T{_C}{^+}$, a curvature from the straight line fit is clearly seen.  This arises due to the presence of  higher order terms close to the transition temperature.  Exponent calculations gives the value of $\gamma{_3}$ to be 4.57, 4.80 and 5.04, for this series of samples, which matches well with the 3-D Heisenberg value of 4.88.\newline Using the scaling relations\newline $\gamma{_3}{^+} = \gamma +2\triangle$ ;\newline $\triangle = \gamma + \beta $;\newline $\delta = 1 + \gamma /\beta $;  and\newline $\alpha +2\beta +\gamma = 2$, \newline (where $\alpha$ ,$\beta$ ,$\gamma$ and $\triangle$ are the exponents associated with the specific heat, magnetisation, susceptibility and the gap exponent\cite{stan} respectively) all the exponents, required for a complete characterisation of the system have been determined, and is given in table 2.  
\begin{table}
\caption{\label{tab:table 2} Values of the all the critical exponents as calculated using linear and nonlinear ac susceptibility in this series of samples. $T{_C}(\chi{_3})$ is the value of the transition temperature determined from the crossover of $\chi{_3}$ in the temperature axis. The values quoted by the previous workers is also given for the sake of comparison} 
\begin{ruledtabular}
\begin{tabular}{cccc}
&A2 &A3 &A4\\ 
\hline 
$\gamma$ &$1.16\pm0.008$ &$1.388\pm0.01$ &$1.41\pm 0.01$\\ 
&1.19$\footnotemark[1]$ &1.38$\footnotemark[1]$ &- \\
\hline 
$\gamma{_3}$ &$4.57\pm0.08$ &$4.805\pm0.12$ &$5.04\pm0.09$\\
\hline
$\triangle$ &1.705 &1.708 &1.815\\
\hline 
$\beta$ &0.545 &0.32 &0.405\\
&0.35$\footnotemark[1]$ &0.41$\footnotemark[1]$ & 0.52$\footnotemark[1]$\\
\hline 
$\alpha$ &-0.25 &0.03 & -0.22 \\
&0.2$\footnotemark[1]$ &-0.2$\footnotemark[1]$ &-0.7$\footnotemark[1]$\\
\hline 
$\delta$ & 3.12 &5.33 &4.48\\
 &4.42$\footnotemark[1]$ &4.49$\footnotemark[1]$ &4.32$\footnotemark[1]$\\
\hline 
$T{_C} (\chi{_3})$ & 267.05 &222.55 &173.45\\
\end{tabular}
\end{ruledtabular}
\footnotetext[1]{From Ref \cite{akm}}
\end{table}  
As is clearly seen our values match reasonably with that given in ref.\cite{akm} The only exponent in which a large difference is seen is $\alpha$ which is calculated by using the Rushbrooke equality, ($\alpha +2\beta +\gamma = 2$).  The ambiguity in the value of this exponent is probably because the values determined in ref\cite{akm} is calculated by using both high dc field and low ac field measurements,  as is the general practice,  whereas in our case $\alpha$ is determined by using low field linear and nonlinear ac susceptibility measurements alone.
\subsection{Determination of $T{_C}$ using Nonlinear susceptibility}
Accurate determination of $T{_C}$ has always been an important consideration in the study of criticality across the ferromagnetic transition.  The most popular techniques used till date have been the K-F formalism  (discussed in the earlier section) and the Arrots plots\cite{arr}.\newline Arrots plots works on the principle of plotting the isothermal experimental data as $M{^2}$ Vs $H/M$ at different temperatures in the transition region.  They should be straight lines in the critical region with the intercepts of these lines with the ($H/M$) axis being positive if  $T> T{_C}$, and negative when $T< T{_C}$.  However, it is important to note that this is valid only when the domain allignment is complete, ie the low field experimental data is completely excluded.  This can at times be a non trivial problem, specially in systems with large field induced effects, causing uncertainties in the accurate determination of $T{_C}$.  Also, in inhomogeneous systems, (like amorphous ferromagnets), the Arrots plots, tend to be curved, even at large measuring fields, thus making the identification of $T{_C}$ even more difficult\cite{aha}.\newline As has been described earlier, $\chi{_3}$ diverges on both sides of the transition temperature, and changes sign exactly at $T{_C}$. Moreover, this critical behaviour is extremely sensitive to the magnitude of the field in which the measurement is done, and this critical feature is seen to vanish with increasing field.  This makes $\chi{_3}$ a very promising candidate for an accurate determination of $T{_C}$, as even by theory, the ferromagnetic transition is defined for the limit $H\Rightarrow 0$, and any large measurement field is likely to change the thermodynamic properties of the system under study.  Plotting the temperature of crossover ($T{_C}$) Vs the field used during the measurement ($H{_{ac}}$), it is clearly seen that the $T{_C}$ varies linearly with the applied field.  Thus, extrapolating this to $H=0$, one will be able to determine $T{_C}$ to a very high accuracy.  This is shown in figure 5, where $T{_C}$ vs H is plotted for the sample A3.  The intercept at $H=0$, gives us the value of $T{_C} = 222.43$K. 
\section{Conclusions}
 A simple molecular treatment  of the second order para-ferromagnetic phase transition shows that $\chi{_3}$ diverges on both sides of the transition, and changes over sign at $T{_C}$. This criticality has been experimentally observed in a well studied series of amorphous alloys, and the related assymptotic critical exponents have been determined.  It is shown that sensitive  nonlinear susceptibility measurements , along with linear susceptibility, can be used to fully characterize the phase transition, without the need for any other type of measurements, like dc magnetisation, specific heat, etc.  Moreover, due to the functional form of $\chi{_3}$ at the transition, the transition temperature $T{_C}$ can be determined accurately.  Using the field dependence of $\chi{_3}$, an extrapolation to true zero field can be made, which gives us a unique way of determining the transition temperature.  Subtle features not seen in the first order susceptibility, like the presence of magnetic clusters with a distribution of $T{_C}$'s is observed.  This only goes on to show that at times nonlinear susceptibility can be a more sensitive tool than their linear counterparts in understanding the magnetism of spin systems in the critical region. 
\section{Acknowledgements} 
We are grateful to Prof. A. K. Majumdar for help, specially in providing us well characterised samples along with all relevant details of earlier measurements.  Acknowledgements are due to Dr. Ashna Bajpai for initiating measurements in the early part of this work, and Mr. Kranti Kumar for help rendered during the course of measurements. 

\newpage
\begin{figure}
\caption{Kouvel Fischer Plots of the real part of the first order susceptibility for the samples A2, A3 and A4 respectively. The inverse of the slope gives the value of the susceptibility exponent $(\gamma)$, and the intercept on the temperature axis provides the transition temperature $(T{_C}$).}
\caption{The critical feature of $\chi{_3}$ as observed in the real part of the third order susceptibility in the samples A2,  A3 and A4 respectively. The measurements were done at a frequency of 133.33 Hz and the exciting field is of the order of 200 mOe.}
\caption{The field dependence of the third order susceptibility as seen in the sample A3 at a measuring frequency of 133.33 Hz. As is clearly seen, the critical behaviour is supressed with increasing applied field.}
\caption{Double logarithmic plots of $\chi{_3}$ versus the reduced temperature for the samples A2, A3 and A4 respectively. The slopes of the straight lines give the value of the exponent $\gamma{_3}$.} 
\caption{The temperature of crossover $T{_C}$ versus the measuring field $H{_a}{_c}$ for the sample A3. A linear behaviour is observed, and the interpolation to zero applied field gives us a unique way of determining the true transition temperature.}
\end{figure} 

\begin{thebibliography}{ }
\bibitem[1]{rang}A. Chakravarti and R. Ranganathan, Solid State Commun. {\bf82}, 591 (1992).
\bibitem[2]{bind}K. Binder and A. P. Young, Rev. of Mod. Phys. {\bf58}, 803, (1986). 
\bibitem[3]{ash1}A. Bajpai and A. Banerjee, Phys. Rev. B. {\bf55},12439,(1997).
\bibitem[4]{ash2}A. Bajpai and A. Banerjee, Phys. Rev. B. {\bf62}, 8996, (2000).
\bibitem[5]{ash3}A. Bajpai and A. Banerjee, J. Phys.:Condens. Matter {\bf13}, 637, (2001).
\bibitem[6]{morin}P. Morin and D. Schmitt, Phys. Rev. B. {\bf23}, 5936 (1981).
\bibitem[7]{bitoh}Teruo Bitoh, Takashi Shirane and Susumu Chikazawa, J. Phys. Soc. Jpn. {\bf62}, 2837 (1993).
\bibitem[8]{sato}Toshikazu Sato and Yoshihito Miyako, J. Phys. Soc. Jpn. {\bf51},1394 (1981).
\bibitem[9]{ram}A. P. Ramirez, P. Coleman, P. Chandra, E. Bruck, A. A. Menovsky, Z. Fisk and E. Bucher, Phys. Rev. Lett. {\bf68}, 2680 (1992).
\bibitem[10]{kaul}S. N. Kaul, A. Hofmann and H. Kronmuller, J. Phys. F {\bf16}, 365 (1986).
\bibitem[11]{akm}A. Das and A. K. Majumdar, Phys. Rev. B. {\bf47}, 5828  (1993).
\bibitem[12]{akm1}A. Das and A. K. Majumdar, J. Magn. Magn. Mater. {\bf128}, 47 (1993). 
\bibitem[13]{akm2} A. Das and A. K. Majumdar, Phys. Rev. B. {\bf43}, 6042 (1991). 
\bibitem[14]{kaul1}S. N. Kaul,  J. Magn. Magn. Mater. {\bf53}, 5 (1985).
\bibitem[15]{chi}S. Chikazawa, H. Arisawa, T. Bitoh, T. Kikuchi, N. Hasegawa and S. Taniguchi, J. Magn. Magn. Mater. {\bf90, 91}, 343 (1990).
\bibitem[16]{ashna}A. Bajpai and A. Banerjee, Rev. Sci. Instrum. {\bf68}, 4075 (1997).
\bibitem[17]{kf}James S. Kouvel and Michael E. Fisher, Phys. Rev. {\bf136}, A1626 (1964).
\bibitem[18]{wada}Koh Wada and Hajime Takayama, Prog. Theor. Phys. {\bf64}, 327 (1980).
\bibitem[19]{fuj}Sumiyoshi Fujiki and Shigetoshi Katsura, Prog. Theor. Phys. {\bf65}, 1130 (1981).
\bibitem[20]{stan}Introduction to Phase Transitions and Critical Phenomena, H. Eugene Stanley, Oxford Science Publications (1971).
\bibitem[21]{arr}Anthony Arrot, Phys. Rev. {\bf108},1394 (1957).
\bibitem[22]{aha}Introduction to the Theory of Ferromagnetism, Amikam Aharony, Clarendon Press. Oxford (1996).
\end{thebibliography}
\end{document}